\documentclass[journal,web]{ieeecolor}
\usepackage{jsen}
\usepackage{cite}
\usepackage{amsmath,amssymb,amsfonts}
\usepackage{algorithmic}
\usepackage{graphicx}
\usepackage{textcomp}
\usepackage{wrapfig}
\usepackage{tabularx}
\usepackage[pagewise, modulo]{lineno}
\usepackage{upgreek}
\usepackage{booktabs}
\usepackage{amssymb}
\usepackage{newtxtext}
\usepackage{newtxmath}
\usepackage{multirow}
\usepackage{tikz}

\DeclareMathAlphabet{\mathfrak}{U}{euf}{m}{n}
\SetMathAlphabet{\mathfrak}{bold}{U}{euf}{b}{n}
\usepackage[zerostyle=e]{newtxtt}
\usepackage{amsmath}
\usepackage{graphicx}

\def\BibTeX{{\rm B\kern-.05em{\sc i\kern-.025em b}\kern-.08em
    T\kern-.1667em\lower.7ex\hbox{E}\kern-.125emX}}
\definecolor{abstractbg}{rgb}{0.89804,0.94510,0.83137}
\newcommand\copyrighttext{%
  \footnotesize \textcopyright \the\year{} IEEE. Personal use of this material is permitted. Permission from IEEE must be obtained for all other uses, including reprinting/republishing this material for advertising or promotional purposes, collecting new collected works for resale or redistribution to servers or lists, or reuse of any copyrighted component of this work in other works.}

\renewcommand\fbox{\fcolorbox{red}{white}}
\newcommand\copyrightnotice{%
\begin{tikzpicture}[remember picture,overlay]
\node[anchor=south,yshift=10pt] at (current page.south) {\fbox{\parbox{\dimexpr0.75\textwidth-\fboxsep-\fboxrule\relax}{\copyrighttext}}};
\end{tikzpicture}%
}

\begin{document}

\title{A multi-channel silicon package for large-scale skipper-CCD experiments}

\author{A. M. Botti, C. Chavez,  M. Sofo-Haro, C. S. Miller, F. Chierchie, M. Jonas, M. Lisovenko, H. Gutti, D. Czaplewski, A. Lathrop, L. Scott, J. Tiffenberg, G. Fernandez-Moroni, J. Estrada
\thanks{A.M Botti and C. Chavez contributed equally to this work.}
\thanks{A. M. Botti is with the Fermi National Accelerator Laboratory, Batavia, IL 60510 USA, and also with the Kavli Institute for Cosmological Physics, University of Chicago, Chicago, IL 60637 USA.}
\thanks{C. Chavez is with Fermi National Accelerator Laboratory, PO Box 500, Batavia IL, 60510, USA., with Universidad Nacional del Sur (UNS), Bah\'ia Blanca, Argentina, and also with Facultad de Ingenier\'ia, Universidad Nacional de Asunci\'on, Asunci\'on, Paraguay}
\thanks{M. Sofo Haro is with Universidad Nacional de C\'ordoba, Instituto de F\'isica Enrique Gaviola (CONICET), and also with Reactor Nuclear RA0 (CNEA), C\'ordoba, Argentina.}
\thanks{C. S. Miller, M. Lisovenko, and D. Czaplewski are with Argonne National Laboratory, 9700 South Cass Ave., Lemont, IL, 60439, USA}
\thanks{F. Chirchie is with Instituto de Investigaciones en Ingeniería El\'ectrica “Alfredo C. Desages” CONICET, Bahía Blanca, Argentina, and also with Universidad Nacional del Sur (UNS), Bah\'ia Blanca, Argentina.}
\thanks{M. Jonas, H. Gutti, A. Lathrop, L. Scott, J. Tiffenberg, G. Fernandez Moroni, and J. Estrada are with Fermi National Accelerator Laboratory, PO Box 500, Batavia IL, 60510, USA.}
\thanks{Corresponding author: A. M. Botti (botti.anamartina@gmail.com)}}


\IEEEtitleabstractindextext{%
\fcolorbox{abstractbg}{abstractbg}{%
\begin{minipage}{\textwidth}%
\begin{wrapfigure}[14]{r}{3in}%
\includegraphics[width=3in]{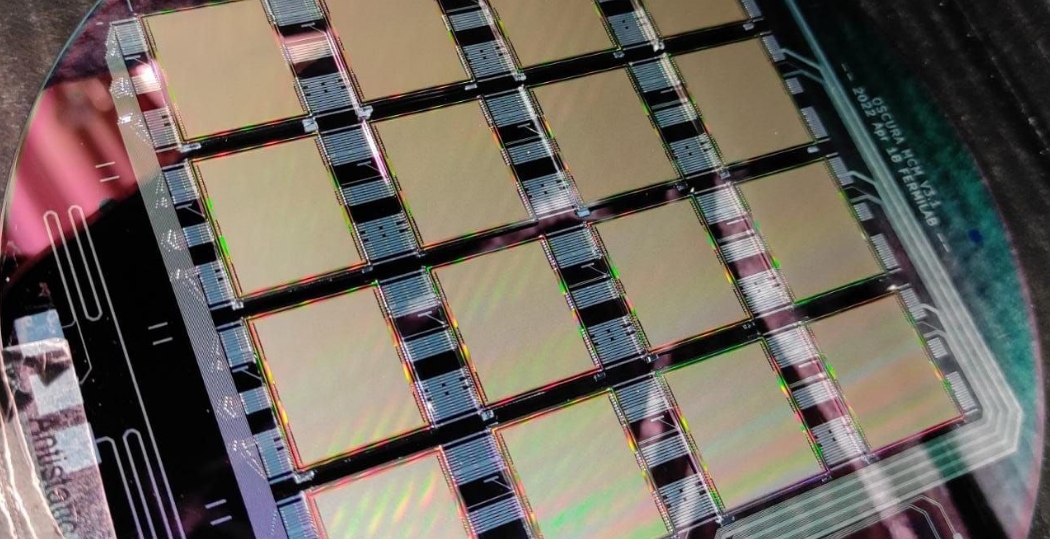}%
\end{wrapfigure}%
\begin{abstract}
The next generation of experiments for light-dark-matter and neutrino searches based on skipper Charge Coupled Devices (skipper-CCDs) introduces new challenges for the sensor packaging and readout architecture. Scaling the active mass while simultaneously reducing the experimental backgrounds in orders of magnitude requires a novel high-density silicon-based package that must be massively produced and tested. In this work, we present a silicon multi-chip module design capable of hosting up to 16 skipper-CCDs, along with the fabrication process for the first prototypes. We thoroughly tested and characterized the assembled prototypes to build an empirical model for the video outputs signal of skipper-CCDs integrated into the silicon package. We then used this model in simulations to optimize the fabrication process and achieve the robust performance required for the full-scale array, which we validated through a new round of prototype fabrication.  We outline the specifications selected for the ongoing production of 1500 silicon wafers that will ultimately add up to a 10 kg skipper-CCD array with 24000 readout channels.
\end{abstract}

\end{minipage}}}

\maketitle
\copyrightnotice

\section{Multi-kilogram skipper-CCD detectors for rare-event searches}

Skipper Charge-Coupled Devices (skipper-CCDs)~\cite{Smith2010, Holland:2003, Janesick1990,Wen1974, Tiffenberg:2017aac} is a new semiconductor-based technology used in a wide range of scientific applications, from imaging to rare-event searches~\cite{Drlica_2020, depaoli2024, barak2023senseisearchmillichargedparticles}. It allows us to measure the number of electron-hole pairs produced after a particle interaction in silicon with deep sub-electron readout noise and, in turn, to detect physical processes with interaction energy as low as the silicon bandgap. To achieve this unprecedented sensitivity, we implement two complementary noise reduction methods. First, to cancel the high-frequency contribution of the noise, we implement a correlated double-sample technique in which the baseline and signals are integrated into a time window and then subtracted~\cite{CDSfiltering}. Then, to reduce the low-frequency contribution, we take multiple non-disruptive measurements of the same charge package and then average these measurements. This approach allows us to decrease the readout noise from a few electrons, as in traditional CCDs, to as low as 0.04 electrons.

Skipper-CCD deep sub-electron resolution opens the door for detecting processes with ${\cal O}$(eV) energy, such as that expected from light-dark-matter interacting electrons in a silicon target~\cite{Crisler:2018gci,Abramoff:2019dfb}. However, to explore the parameter space region of most interest given the cosmological milestone scenarios, we need a total exposure of about 30\,kg\,year; efforts to scale skipper-CCD detectors in mass are ongoing within the SENSEI (100\,g), DAMIC-M (1\,kg), and OSCURA experiments (10\,kg)~\cite{senseicollaboration2023sensei,DAMIC-M:2023gxo,aguilar2022oscura}. 


The transition to a multi-kilogram detector poses new challenges for skipper-CCDs, mainly due to a two-order-of-magnitude increase in the number of readout channels and data volume compared to current instruments. This shift necessitates a new paradigm for readout electronics, sensor packaging, infrastructure, and processing tools. Additionally, the OSCURA design goals require a two-order-of-magnitude reduction in radiogenic backgrounds, which calls for innovative methods in the production, storage, and materials for the sensor packaging. To address these challenges, we developed a novel high-density packaging solution that enables low background, synchronized readout, and uniform operating conditions for the 24,000 CCDs in the 10\,kg detector. 

In this work, we introduce the basic unit of this package, which consists of an aluminum circuit on a silicon substrate capable of accommodating up to 16 devices. In section~\ref{sec:design}, we discuss the requirements for the design, while in section~\ref{sec:fabrication}, we detail the fabrication and assembly process of the silicon-based package. In sections~\ref{sec:testing} and ~\ref{sec:results}, we present the warm and cold electrical tests and the CCD signal characterization, and in section~\ref{sec:sim}, the signal model that allowed us to define modifications in the design parameters discussed in section~\ref{sec:newDesign}. We conclude with a summary in~\ref{sec:outlook}.

\section{The OSCURA package}
\label{sec:design}

The OSCURA targeted sensitivity is only reachable if we understand and mitigate the background contribution of the radioactive material close to the sensor, as in the electronic components, flex cables, and mechanical support pieces. Furthermore, to prevent dark-current single electrons from accumulating into multiple-electron events that contribute to the signal background, we require a maximum readout time of a few hours, achievable by increasing the number of readout channels. 

The CCD size and form factor are trade-offs defined by the fabrication yield criteria and the requirements previously described: a bigger sensor with a higher active volume and number of channels will be more likely to have a fabrication defect. The OSCURA CCDs, fabricated at Microchip Technology Inc with a yield higher than 90\%, have 1278\,$\times$\,1058 pixels, with a pixel size of (15\,$\upmu$m\,$\times$\,15\,$\upmu$m\,$\times$\,675\,$\upmu$m). Given this geometry, 10\,kg translates to 24000 CCDs or 28\,Gigapixels. We devised the CCD package to facilitate this scale, using one channel per CCD and a total readout time of 2 hours with a noise of about 0.15$\sim$0.20\,e$^-$ and a background goal of 0.01\,events/kg/day/keV. We thoroughly discuss the OSCURA requirements in~\cite{oscurasensors}.

\begin{figure}[t]
    \centering
    \includegraphics[width=1.00\linewidth]{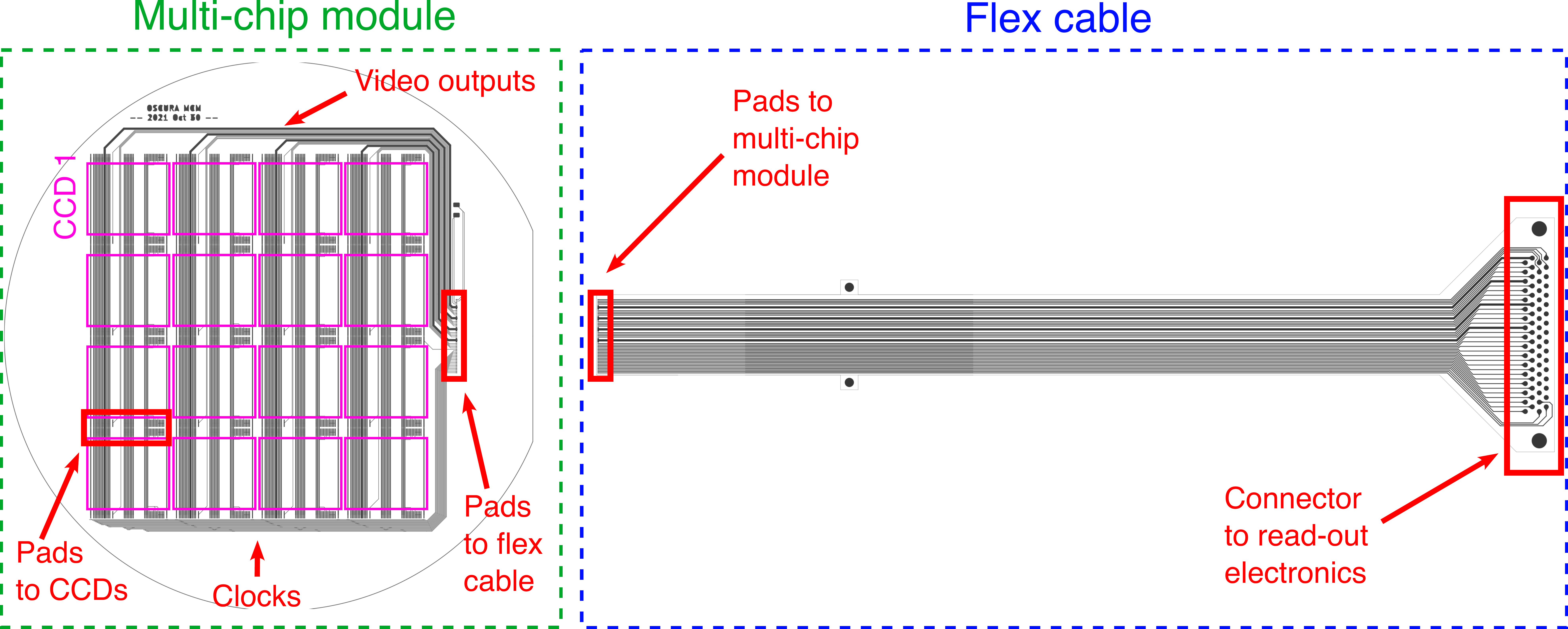}
    \caption{Design layout for a 16 channel CCD package. (Dashed-green box) 6'' wafer layout to map the CCD pads to the readout flex cable. (Dashed-blue box) flex cable to connect the CCD pads to the readout electronics. The cable design is cropped to fit the figure.}
    \label{fig:designLayout}
\end{figure}

We present the design of the basic unit of the OSCURA detector in Fig.~\ref{fig:designLayout}, a.k.a. the {\it{multi-chip module}}. It consists of a 6'' substrate with traces (green-dashed box) that connect up to 16 CCD (pink squares) to the readout electronics via a flex cable (blue-dashed box). We arranged the traces for all CCD video outputs and transistor's supply voltage on one side of the wafer while we placed the shared time-varying operating voltages, or clocks, and bias voltages on the opposite side. The pads for connecting the flex cable are positioned as far as possible from the CCDs to minimize the risk of radioactive decays in the Kapton flex circuit reaching the sensors~\cite{richardFlex} and leaving space to glue the cable in the right side of the design.

The OSCURA array consists of 1500 multi-chip modules, grouped into super modules of 16 each. We process each super module with one analog-to-digital (ADC) converter channel of a low-threshold acquisition board~\cite{cancelo2021low}. To this aim, we implement two stages of analog multiplexed electronics: one between the multi-chip and the super modules and one between the super module and the low-threshold acquisition board~\cite{chavez2022_multiplexed}. The first multiplexing stage switches the 16 CCD video analog outputs from one multi-chip module, and the second stage switches between the 16 outputs from the first stage into the ADC channel.

To validate the electronics and establish a benchmark for the package performance, we utilized multi-chip modules fabricated on a ceramic substrate~\cite{chavez2022_multiplexed}. However, the substrate holds the most abundant material in the volume surrounding the sensors, and the radioactive materials in the ceramic can significantly contribute to the background. Therefore, we must fabricate the multi-chip modules on an ultra-low-background material, like the silicon used in CCDs.

\section{Multi-chip-module fabrication and assembly}
\label{sec:fabrication}

We fabricated the silicon multi-chip module prototypes at the 
Center for Nanoscale Materials at Argonne National Laboratory (ANL) through a user proposal; the definition of the fabrication parameters and recipes is a joint effort between Fermilab and ANL.

\begin{figure}[t]
    \centering
    \includegraphics[width=0.7\linewidth]{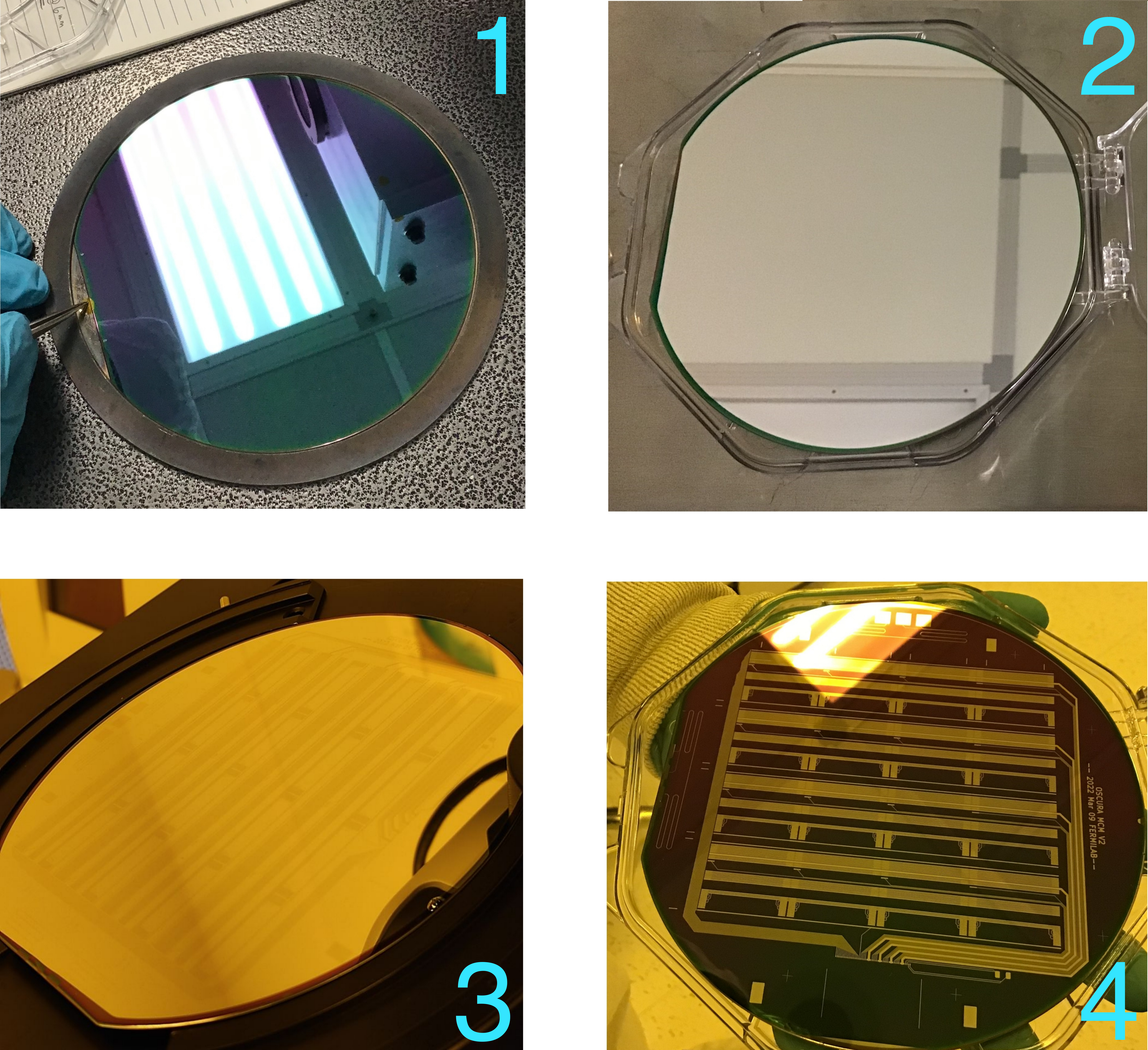}
    \caption{Pictures of the silicon wafer after fabrication step. (1)~SiO$_2$ deposition. (2)~Aluminum deposition. (3)~lithography and development. (4)~aluminum etching.}
    \label{fig:mcmProd}
\end{figure}

We fabricated the first silicon package using a 6'', high-resistivity, $\langle100\rangle$ n-type silicon wafer with a thickness of 675~$\upmu$m , the same type as in CCDs.
To isolate the silicon from the metal, we deposit a 390\, nm silicon dioxide (SiO$_{2}$) layer with about 4\% uniformity through a plasma-enhanced chemical vapor deposition with an Applied Materials (AMAT) P5000 operated at 400$^\circ$C. Panel 1 in Fig.~\ref{fig:mcmProd} shows a picture of the wafer after the SiO$_{2}$ deposition. In addition, we also tried 6'' low-resistivity $\langle100\rangle$ p$^{++}$-doped silicon wafers with a thermal growth SiO$_2$ layer of 3\,$\upmu$m. We used p$^{++}$-doped silicon because we had immediate access to these wafers.

We then used alternatively an AJA Metal sputtering system or an angstrom quantum sputtering system to deposit a 450\,nm Aluminum layer on top of the SiO$_{2}$. We show a picture of the wafer after the sputtering on panel 2 in Fig.~\ref{fig:mcmProd}. After baking the wafer for 5 minutes at 130$^\circ$ in a YES oven to release the moisture, we spun a 1 $\upmu$m layer of SPR-955 photoresist and baked it for 2 minutes at 90\,C$^\circ$ on a hotplate. We then used a Heidelberg MLA 150 Maskless Lithography tool to expose the photoresist according to the design presented in section~\ref{sec:design}. The exposure took about 30 minutes with a 50-second development using Microposit$^\mathrm{TM}$ MF$^\mathrm{TM}$-CD-26 developer. In panel 3 of Fig.~\ref{fig:mcmProd}, we present a picture of the wafer with the patterned photoresist right after the development. 

Next, we wet etched the wafer using aluminum etchant type A at 40$^\circ$C until the tracks became visible (about 2.5 minutes). As we display in panel 4 of Fig.~\ref{fig:mcmProd}, the photoresists successfully protected the aluminum traces while the etchant removed the rest of the metal. We finally cleaned the remaining photoresists with Microposit$^\mathrm{TM}$ Remover 1165.

The second layer consists of a SiO$_2$ coating that works as passivation to protect the aluminum tracks from scratches and prevent shorts during and after the assembly. This layer needs openings at the pad positions to connect the CCDs. To this aim, we deposit a 490\,nm SiO$_2$ layer using the AMAT P5000 system operated at 275$^\circ$C. We performed an HMDS treatment with a YES-TA oven to help the photoresist adhere to the SiO$_2$ and spun 1 $\upmu$m of SPR-955. We proceeded with the lithography and development in the same way as for the aluminum layer. We then used a dry etching Oxford Plasmalab 100 tool to remove the SiO$_2$ from the pads (about 25 minutes), after which we removed the remaining photoresist with 1165. Finally, we used an e3511 ESI Plasma Asher to remove any residue.

\begin{figure}[t]
    \centering
    \includegraphics[width=1.00\linewidth]{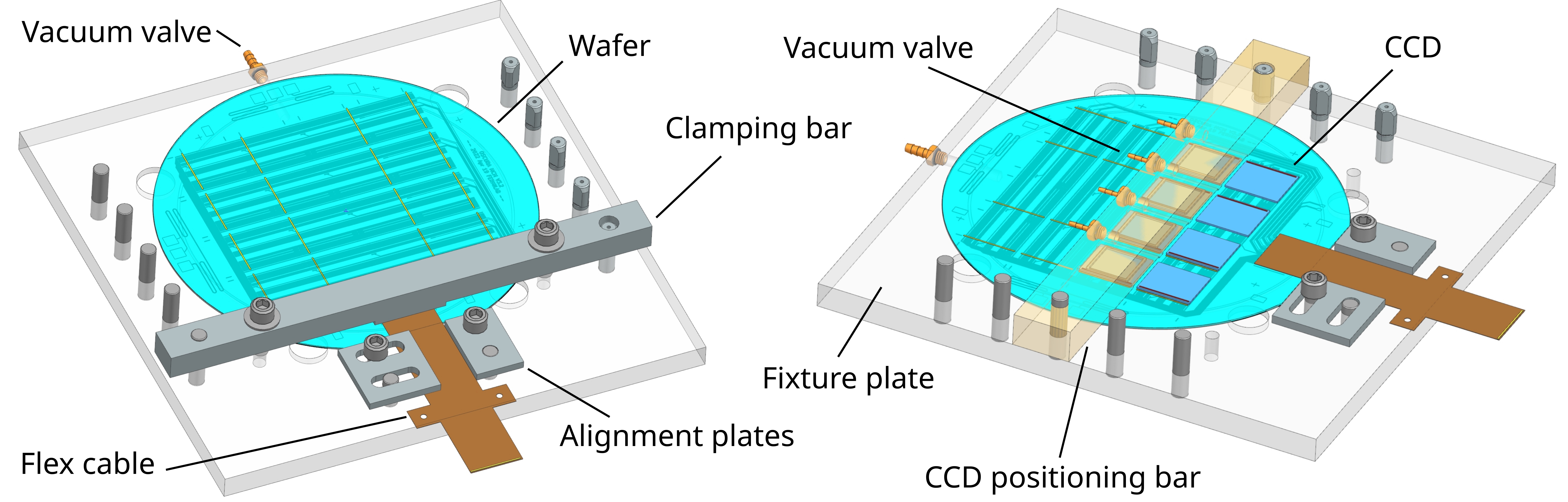}
    \caption{Schematics of the tools used to assembly a multi-chip module. (Left) tools used to glue the flex cable. (Right) tools used to glue the 16 CCDs.}
    \label{fig:mcmAssembly}
\end{figure}

We performed the assembly process at Fermilab. The first step is to attach the flex cable to the silicon circuit that connects the silicon substrate pads to the readout electronics. To achieve this, we designed a fixture that holds the silicon wafer on a tray and incorporates a series of alignment tools. We show a design view of these tools in the left panel of Fig.~\ref{fig:mcmAssembly}. The fixture plate consists of a hollow aluminum tray with a pocket to place the wafer and a valve on one side. The hole in the tray pocket fixes the wafer by connecting a vacuum pump so it does not move during assembly. The flex cable is then manually placed using the aligning plates on the side of the tray to match it with the multi-chip module pads. We apply Pyralux\textsuperscript{\textregistered} laminate at the end of the flex and hold it with the clamping bar to apply uniform pressure. Then, we disconnect the vacuum pump and put the entire assembly setup inside an oven at 190$^\circ$C for 3.5 hours. 

The second step in the assembly process consists of gluing the CCDs onto the silicon substrate. To this end, we designed the tool in the right panel of Fig.~\ref{fig:mcmAssembly} that allows the simultaneous placement of up to four CCDs. The CCD positioning tool is a hollow bar with four pockets and four valves to pick and place each CCD. To prevent the glue from flowing onto the pads, we use double-sided Kapton\textsuperscript{\textregistered} tape, which we place on each CCD edge. 

\begin{figure}[t]
    \centering
    \includegraphics[width=1.00\linewidth]{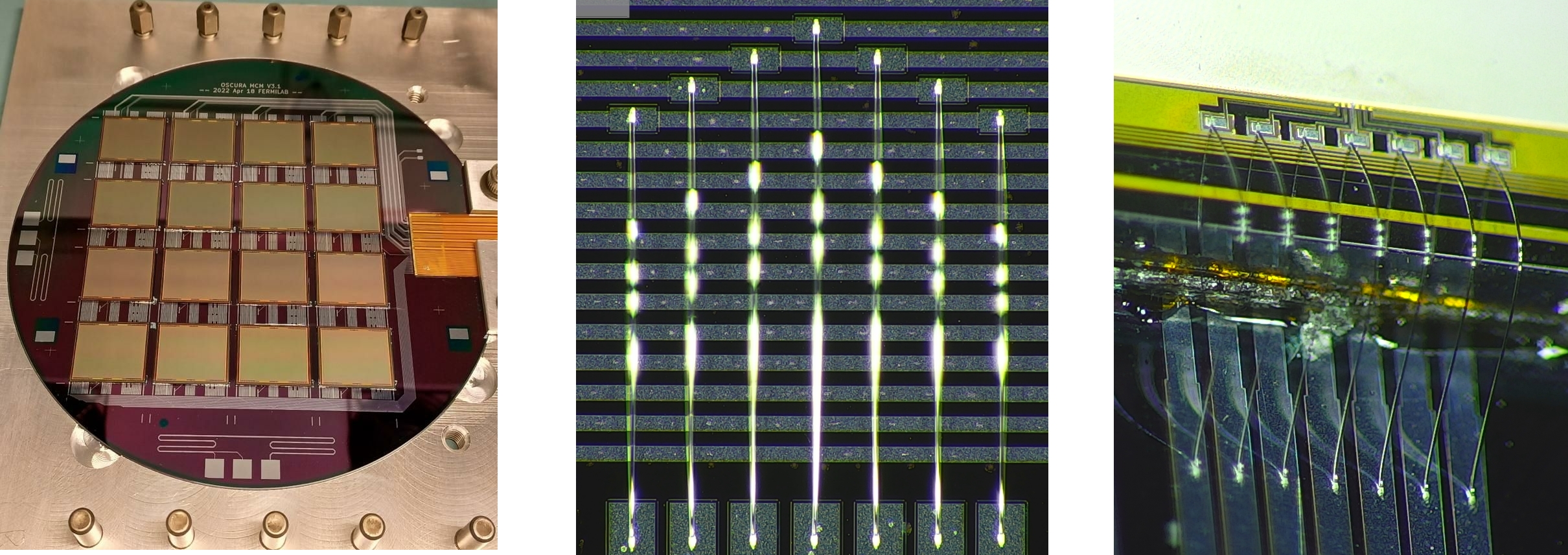}
    \caption{Pictures of a multi-chip module after assembly. (Left) silicon wafer with 16 CCDs and flex cable, still on the assembly tray. (Middle) aluminum wire-bonds between padrs on the silicon package; the SiO$_2$ openings for pad connection can be seen as a thin line around the pads. (Right) wire bonds between a CCD and the silicon wafer.}
    \label{fig:finishedMCM}
\end{figure}

As in the previous step, we hold the silicon substrate on the fixture plate with vacuum. Then, we position the CCD bar on top of the wafer using one set of alignment pins, as illustrated in the right panel of Fig.~\ref{fig:mcmAssembly}. 
Once the positioning bar is aligned, we lowered the CCDs to make contact with a double-sided tape, which both holds them in place and acts as a spacer, creating a gap that we will later fill with liquid epoxy. After the CCDs are in contact with the tape, we release the vacuum and remove the positioning bar. We repeat this procedure for all four columns until the 16 CCDs are attached to the wafer with double-sided tape. We then apply Epoxy (Master Bond EP30-2) under each of the CCDs mid-way between the two tape locations, which takes 36 hours to cure at room temperature. We display the package in this state, which consists of the silicon substrate with the flex cable and 16 glued CCDs, in the left panel of Fig.~\ref{fig:finishedMCM}.

Finally, we use a Hesse Mechatronics BJ820 fully automatic fine-wire wedge bonder to connect the pads both inside the multi-chip module and between the module and CCDs using 25\,$\upmu$m aluminum wires. We show a picture of connected pads on the multi-chip module in the middle panel of Fig.~\ref{fig:finishedMCM}. The passivation layer on the silicon substrate is well aligned with the aluminum layer, as the pad openings appear as a thin line around the pad. For completeness, we show in the right panel of Fig.~\ref{fig:finishedMCM} wire bonds between the CCD pads and the silicon substrate with the Epoxy overflowing at the CCD edge. Note that this procedure is robust against minor overflows. 

For reference, we present a summary of the wafer fabrication and assembly process in Table~\ref{table:process}, where we indicate the tools and materials used in each step together with a short description.

\begin{nolinenumbers}
\begin{table}[t!]
\renewcommand{\arraystretch}{1.2}
\begin{tabular}{l l l} 
\toprule

\textbf{Step} & \textbf{Description} & \textbf{Tool or material}\\
\midrule 

\multicolumn{3}{c}{\multirow{2}{*}{\textbf{Fabrication}}}\\\\
1 & SiO$_2$ first layer (n-type wafer) & AMAT P5000\\
\multirow{ 2}{*}{2} & \multirow{ 2}{*}{Aluminum deposition } & AJA Metal sputtering or \\
& & Angstrom quantum sputtering \\
3 & Soft bake to remove moisture & YES Oven\\
4 & SPR955 photoresist layer & Spinner\\
5 & Photoresist bake & Hot plate\\
6 & Photoresist exposure & Heidelberg MLA 150\\
7 & Photoresist development & Microposit$^\mathrm{TM}$ MF$^\mathrm{TM}$-CD-26\\
8 & Wet etch  & Aluminum etchant type A\\
9 & Photoresist clean & Microposit$^\mathrm{TM}$ Remover 1165\\
10 & SiO$_2$ second layer & AMAT P5000 \\
11 & HMDS treatment & YES-TA Oven \\
12-15 & \multicolumn{2}{r}{Repeat steps 4 to 7} \\
16 & Dry etch & Oxford Plasmalab 100\\
17 & \multicolumn{2}{r}{Repeat step 9} \\
18 & Photoresist descum & e3511 ESI Plasma Asher\\
\midrule
\multicolumn{3}{c}{\multirow{2}{*}{\textbf{Assembly}}}\\\\
19 & Flex cable placement & Fixture plate\\
20 & Flex cable glue & Pyralux\textsuperscript{\textregistered} laminate\\
21 & Glue bake & 190$^\circ$C Oven\\
22 & CCD placement & Positioning bar\\
23 & CCD glue & Epoxy Master Bond EP30-2\\
24 & Wire-bonding & Hesse Mechatronics BJ820\\
\bottomrule
\end{tabular}
\caption{Process summary for multi-chip module fabrication and assembly; we list the steps in chronological order. We show pictures taken after steps 1, 2, 7, and 9 in panels 1, 2, 3, and 4 of Fig.~\ref{fig:mcmProd}. The left panel of Fig.~\ref{fig:mcmAssembly} shows the setup during steps 19 to 21 and the right panel from steps 22 and 23. The left panel of Fig.~\ref{fig:finishedMCM} shows the module after step 23, while the middle and right panels provide close-ups of two areas in the MCM where wire-bonds pads are visible after step 24.
}
\label{table:process}
\end{table}
\end{nolinenumbers}

\section{Testing}
\label{sec:testing}

Before biasing the CCDs, we look for short circuits in the package to avoid damage to the sensors or electronics. We perform warm electrical tests to detect fabrication defects in the multi-chip module and flex cable by measuring the resistance between all pads and comparing the results to values in a template of expected ranges. We built these by measuring the resistance between lines of well-performing sensors and extrapolating them to the trace length of the multi-chip module. A low resistance between two or more lines reveals short circuits in the substrate, the wire bonds, or the sensors. 
 
In the second warm test, we connect the multi-chip module to the readout system to check for alterations to the nominal voltage set. Defective sensors will produce a voltage drop when applying the bias voltage to the 16 CCDs. Any variation in the bias translates into a faulty sensor that will affect the other 15 since they share the control signals and reverse bias voltage.

\begin{figure}[t]
\centering
\includegraphics[width=1.0\linewidth]{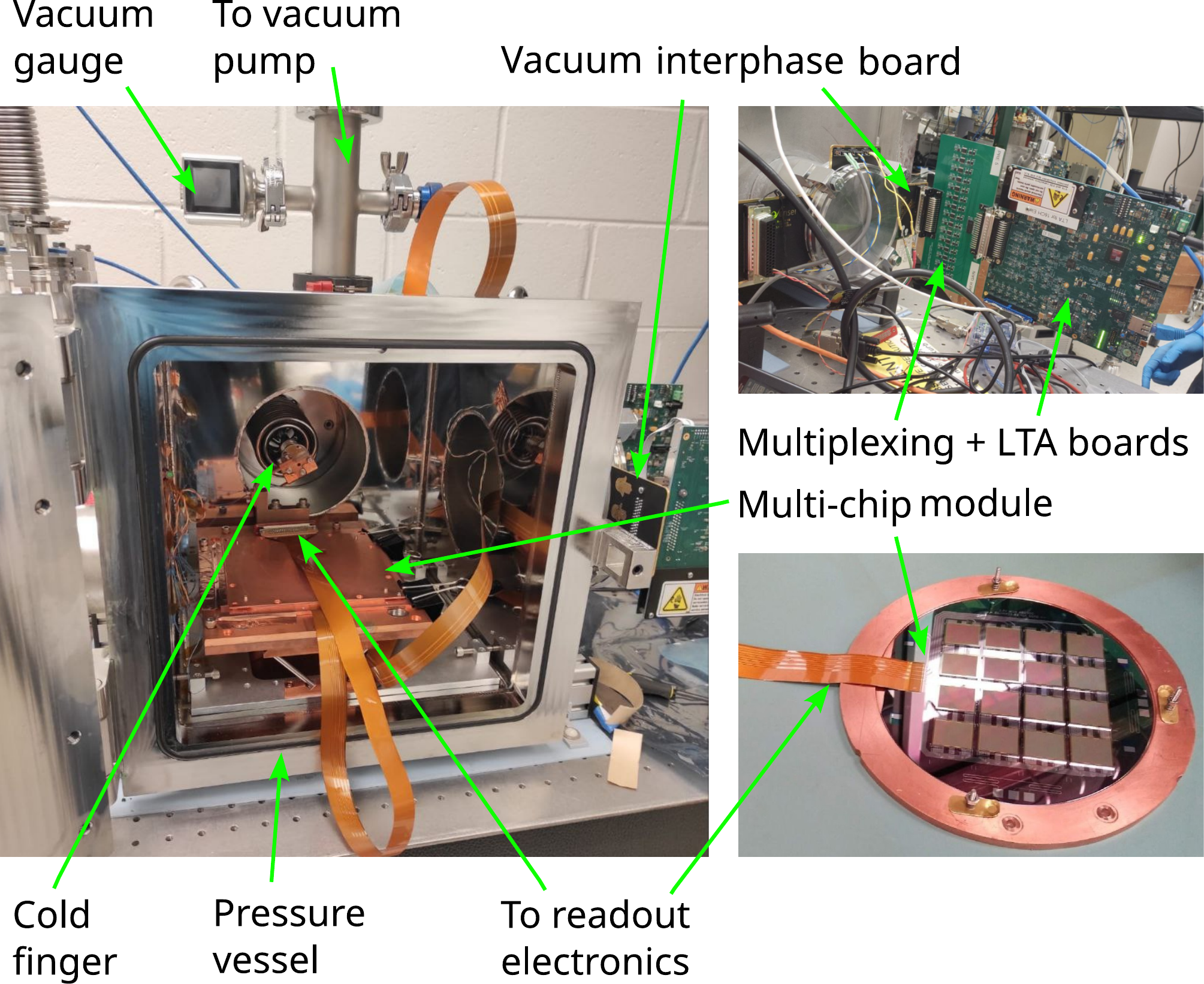}
\caption{Testing setup for silicon multi-chip module. The module is deployed in a copper tray in contact with a cold finger connected to a cryocooler. We use a heater with a PID temperature controller to run the tests at about 160\,K. The electronics consist of an LTA and multiplexing boards.}
\label{fig:test-vessel}
\end{figure}

After testing for defects, we perform cold operation tests and measurements to characterize the CCD video output signals. We deploy the multi-chip modules into the system illustrated in Fig~\ref{fig:test-vessel}, which consists of a vacuum vessel with a supporting pump that maintains a vacuum below 2$\times10^{-4}$\,torr, a cryocooler to cool down the sensors through the cold finger indicated in the picture, and a PID temperature controller to guarantee an operating temperature of 160\,K. The readout electronics consisted of a multiplexing prototype~\cite{chavez2022_multiplexed}, with 16 analog correlated double sampling channels~\cite{Sofo_Haro_2021_Analog_P}, and a low-threshold acquisition board~\cite{cancelo2021low} connected to the multi-chip module through a vacuum interface board.
        
\section{Results for first prototypes}
\label{sec:results}

The first four wafers showed multiple scratches and short circuits between traces caused by incomplete aluminum etching and did not pass the warm test. The electrical test of the fifth wafer showed that the resistance of the aluminum traces and capacitance between traces are significantly higher than in the ceramic prototype~\cite{chavez2022_multiplexed}. In addition, during the nominal voltage test of the n-type wafer, we found that the current-voltage ratio in the bias voltage line followed a nonlinear behavior. We attribute this behavior to an unintentional n-type-silicon-aluminum non-ohmic contact similar to a diode, created during wire-bonding: since the $\mathrm{SiO_2}$ layer between the silicon and the aluminum is < 1\,$\upmu$m, the impedance of this contact depends on the applied voltage~\cite{NORTHROP1971557,1478463}. Furthermore, the non-negligible conductivity of the silicon substrate produces a significant coupling between the clock and video output tracks.

\begin{figure}[t]
    \centering
    \includegraphics[width=1.0\linewidth]{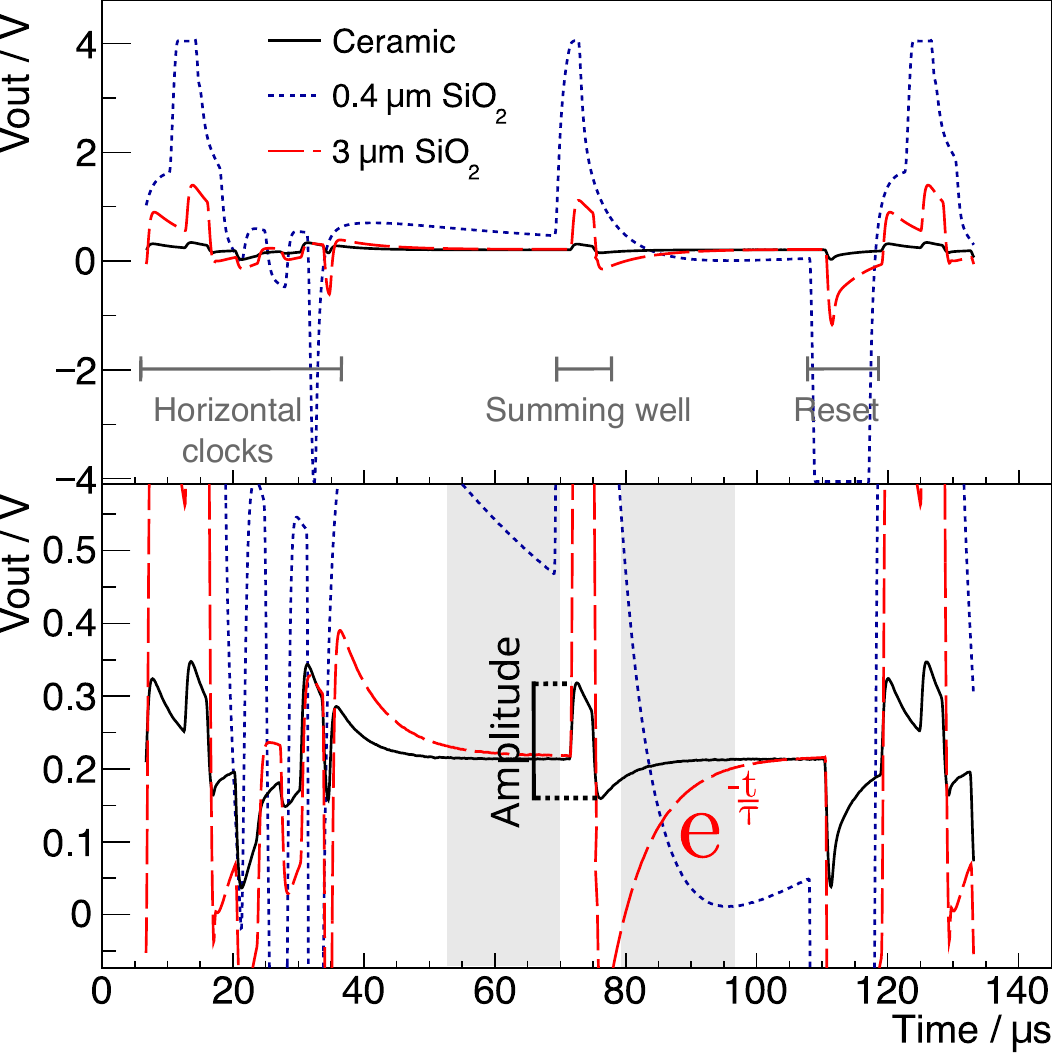}
    \caption{Video output signal for ceramic and silicon multi-chip modules with no CCDs connected. The signal shape is the result of the coupling from different clocks. Both panels show the same data with different y-axis scales.}
    \label{fig:signal-distortion}
\end{figure}

We present in Fig.~\ref{fig:signal-distortion} the video output signals in multi-chip modules fabricated with three different substrates: ceramic (solid black), high-resistivity n-type silicon with a 0.4\,$\upmu$m SiO$_\mathrm{2}$ layer between silicon and aluminum (dotted blue), and low-resistivity p$^{++}$ silicon with a 3\,$\upmu$m (dashed red) SiO$_\mathrm{2}$ layer. Both panels show the same data with different scales obtained using modules with no CCDs connected, so the signals result only from the coupling between clocks and the video output. The peaks at about 20 $\upmu$s correspond to the horizontal clocks that move the charge through the readout stage, at 70 $\upmu$s to the so-called summing well clock, the gate used to move the data to the sensing node, and at about 110 $\upmu$s the reset gate~\cite{Tiffenberg:2017aac}. We calculate the signal charge with the analog correlated double sample technique, in which we first integrate the baseline (gray shadow before the summing well), then move the charge to the sensing node by clocking the summing well, and integrate the signal (gray shadow after the summing well). The difference between the integration before and after the summing well gives the value for one analog skipper sample (ANSAMP). The parameters relevant to this work are the decay time after the summing well ($\tau$ in the figure) and the signal coupling amplitude.

\begin{figure}[t]
    \centering
    \includegraphics[width=1.0\linewidth]{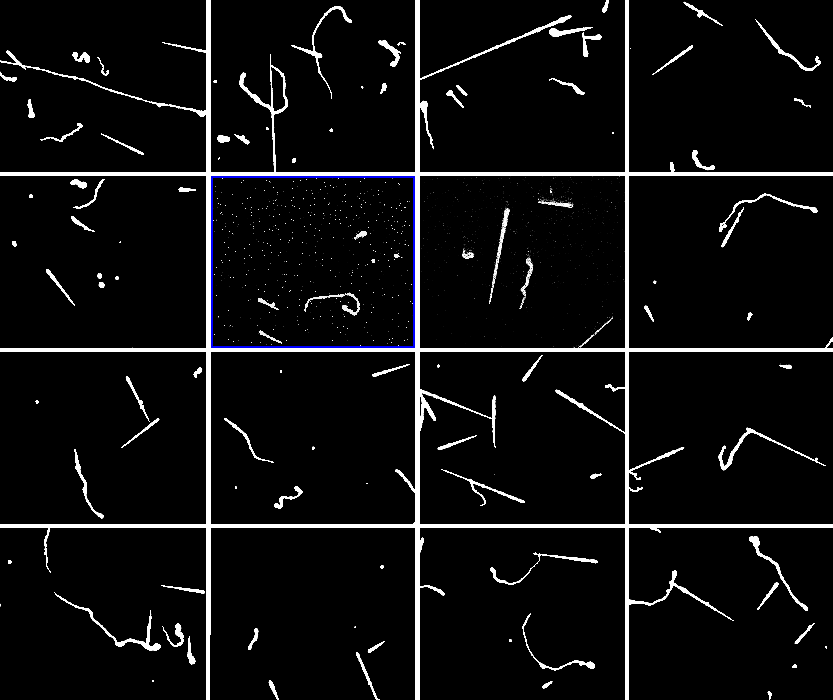}
    \caption{Example image obtained with the multi-chip module built on p$^{++}$-type silicon and a 3\,$\upmu$m SiO$_2$ layer. Each panel corresponds to a CCD segment. In general, straight-like traces are produced by muons, curly-like by electrons, and point-like by photons.}
    \label{fig:mcm-first-img}
\end{figure}

In the first silicon wafer, with n-type silicon and 0.4\,$\upmu$m of SiO$_2$, the output signal is dominated by the coupling of the clocks: the signal amplitude and recovery time are orders of magnitude higher than in the ceramic version and do not allow us to determine the charge in the pixel since it saturates the electronics. However, for the second multi-chip module, with p$^{++}$-type silicon and a 3\,$\upmu$m SiO$_2$ layer, the coupling is at least four times weaker, which allowed us to obtain the first images after connecting the CCDs, as we present in Fig.~\ref{fig:mcm-first-img}, where each of the panels corresponds to one CCD in the module. Straight and curly traces correspond to muons and electrons interacting in the silicon bulk. 

\begin{figure}[t]
    \centering
    \includegraphics[width=1.0\linewidth]{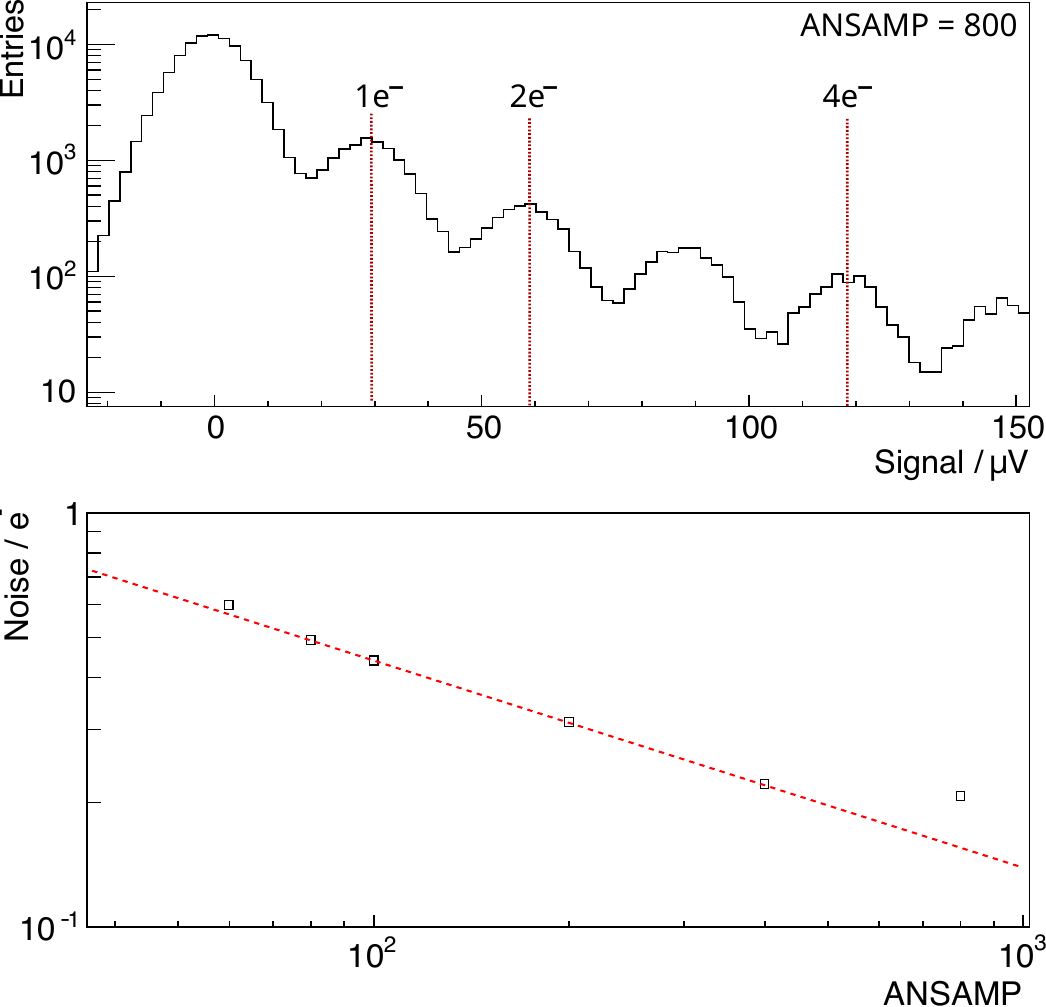}
    \caption{(Top) Pixel charge histogram after subtracting the baseline with 800 analog samples (ANSAMP). (Bottom) noise as a function of ANSAMP; the noise performance is not affected by the distortions.}
    \label{fig:noise}
\end{figure}

Even though the coupling is much larger than in the ceramic version, we obtain single electron resolution as presented in the top panel of Fig.~\ref{fig:noise}. The histogram contains the pixels charge distribution after averaging 800 analog skipper samples, which resulted in a readout noise of about 0.21\,e$^-$, extracted by fitting the 0-electron peak with a Gaussian. Each peak corresponds to an integer number of electrons, where the 1-electron signal, in the top panel of Fig.~\ref{fig:noise}, has a mean value of 30\,$\upmu$V (after converting from digital units). In the bottom panel of Fig.~\ref{fig:noise}, we present the readout noise as a function of the number of analog samples. The red line corresponds to the expected noise behaviour of skipper-CCDs~\cite{Tiffenberg:2017aac} as a function of the number of samples: $\sigma_N = \sigma_\mathrm{N=1}/\sqrt{\textit{N}}$, where $\sigma_\mathrm{N=1} = 4.4$ is calculated from the noise at 100 ANSAMP. The noise follows the expectation in the range between 80 and 500 ANSAMP. For ANSAMP < 80, the signal gain is outside the dynamic range of the integrator capacitor of the pile-up circuit~\cite{Sofo_Haro_2021_Analog_P}, and its noise dominates the output. For ANSAMP = 800, the correlated noise from the testing system produces a deviation from the $\sigma_N = \sigma_\mathrm{N=1}/\sqrt{\textit{N}}$ expectation. We corroborated that this noise level is related to the testing system and not to the package itself by deploying the silicon module in the system used to test the ceramic wafers, where we achieved a noise of about 0.17\,e$^-$ with 400\,ANSAMP.

\begin{figure}[t]
    \centering
    \includegraphics[width=1.0\linewidth]{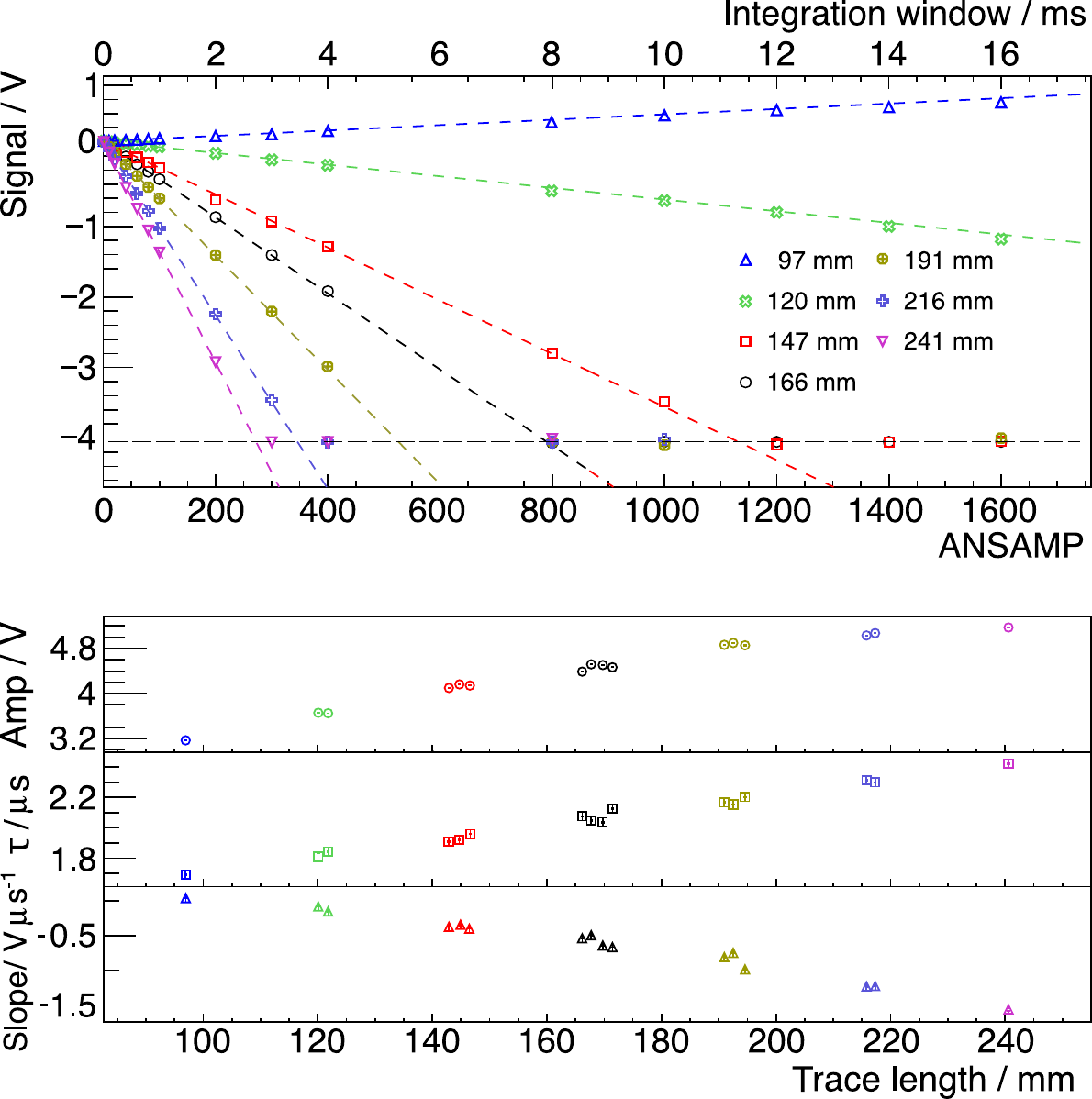}
    \caption{(Top) Signal output of empty pixels for different video channels in the same multi-chip module as a function of ANSAMP. (Bottom) Signal amplitude, time constant and slope extracted from the top panel as a function of the video output trace length. Colors match  approximately the trace length according to the legend in the top panel.}
    \label{fig:signalSat}
\end{figure}

Due to the signal coupling in the multi-chip module, the recovery time after the summing-well clock produces an offset between the baseline and the signal. Since the integrator capacitor accumulates each sample during the analog correlated double sample, the amplitude of the output signal depends on the number of samples. Despite having no electrons in the pixel, each sample adds the offset value to this output signal, which will eventually saturate the ADC in the readout electronics. In the top panel of Fig.~\ref{fig:signalSat}, we present the mean empty pixel signal as a function of the number of analog samples (and integration time) for different CCDs (different markers-colors) in the multi-chip module. Dashed lines correspond to the extrapolation of a linear fit using the data obtained with ANSAMP < 100, and the black dashed line at about -4\,V to the saturation point of the ADC. The legend indicates the length of the video output traces. The offset introduced in each sample is different depending on this length since the coupling between the clocks and output signals, and in turn, the recovery time after the summing well clocking varies according to the electrical properties of the line. For the longest trace (pink-inverted-triangles), the ADC saturates at about 200 ANSAMP; for the shortest trace (blue triangles), the negative offset from the coupling is compensated with a fixed positive offset in the electronics.   

In the bottom panel of Fig.~\ref{fig:signalSat}, we present the video output amplitude (top) and the recovery time $\tau$ (middle) product of the summing well gate coupling, as indicated in the bottom panel of Fig~\ref{fig:signal-distortion}, as a function of the trace length. In addition, we present the slope (bottom) extracted from Fig.~\ref{fig:signalSat} top corresponding to the offset introduced every $1$\,ms of integration. The coupling amplitude and $\tau$ group according to the length of the video output trace: the longer the trace, the stronger the coupling between clocks and video output. It is worth mentioning that in Fig.~\ref{fig:noise} we show data for a channel with a 143\,mm trace length, which does not saturate at ANSAMP=800.

\section{Signal model}
\label{sec:sim}

\begin{figure}[t]
    \centering
    \includegraphics[width=0.9\linewidth]{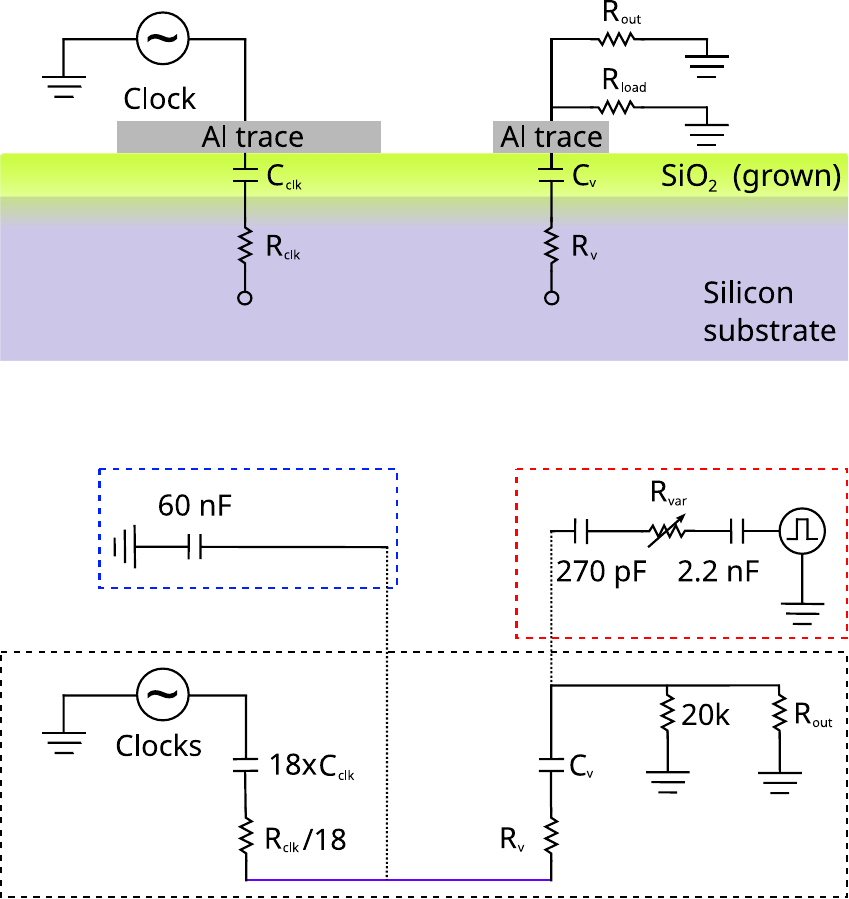}
    \caption{(Top) Diagram of the multi-chip module layer interfaces. (Bottom) A simplified circuit model of the mesh formed between the video output and clocks through the substrate (dashed-black box). It also includes the schematics for the setup to measure $\mathrm{R_{v}}$ and $\mathrm{R_{clk}}$ (red-dashed box) and the modifications introduced in the new design (blue-dashed box).}
    \label{fig:substrate}
\end{figure}

We developed an empirical signal model that reproduces the video output signals after the clocking to assess the impact of the multi-chip module electrical properties on signal distortion. In the top panel of Fig.~\ref{fig:substrate}, we present a simplified schematic of the silicon substrate with a clock and video output trace coupling through the silicon substrate. This multi-chip module consists of an aluminum-dioxide-silicon layer structure with p$^{++}$-type silicon, as in a metal-oxide-semiconductor (MOS) capacitor where the semiconductor node is floating. For the operating voltages of the CCDs, these MOS capacitors are reverse-biased with a high voltage, which fixes their operation point to an accumulation regime~\cite{hu2010modern}. Therefore, $\mathrm{C_{clk} = 2.3\,nF}$ and $\mathrm{C_v= (111 \sim 276)\,pF}$ represent the oxide capacitance between a clock or video-out trace and the silicon substrate, which are constant and determined by \[\mathrm{C} = \frac{\epsilon_0 \epsilon_{\mathrm{ox}}}{\mathrm{d}} \mathrm{L w}  =  11 \mathrm{\frac{pF}{mm^2} A},\] where $\epsilon_0$ is the vacuum permittivity, $\epsilon_{ox}$ the SiO$_2$ dielectric constant, $L$ the trace length (90$\sim$250 mm), w the width (0.1 $\sim$ 0.8\,mm), $\mathrm{A = Lw}$ the area covered by the trace on the substrate, and $\mathrm{d}$ the thickness of the $\mathrm{SiO_2}$ layer. All clock traces have the same area and thus the same capacitance to the substrate ($\mathrm{C_{clk}}$). 

$\mathrm{R_{load}=20 k\Omega}$ is the load resistance at the input of the electronics, while $\mathrm{R_{clk}\ and\ R_v}$ represents the resistance of the clock or video output trace and the silicon substrate determined by $\mathrm{R_{clk} = \frac{R_a}{A_{clk}}}$ and $\mathrm{R_{v} = \frac{R_a}{A_{v}}}$. We define $\mathrm{A_{clk,v} = Lw}$ as the trace area, while $\mathrm{R_a}$ is the film resistance with units of $\mathrm{\Omega\ mm^2}$ which we expect to depend on the $\mathrm{SiO_2-Si}$ interface composition and structure. $\mathrm{R_{out}}$ is the output resistance of the CCD readout transistor, operated as a follower, and depends on its polarization point. 

Following Thevenin's theorem, we can write the equivalent electrical circuit by modeling the 18 parallel clock lines as one, as presented in the bottom panel of Fig.~\ref{fig:substrate} (black dashed box). Then, the recovery time of the coupling at the video output is \[\tau = \mathrm{RC = (R_{out} \parallel 20 k \Omega+R_v+R_{clk}/18) \times (18 C_{clk}\,\text{-{}-}\, C_v)},\]

where $\mathrm{R_{out} \parallel 20 k \Omega = (R_{out}^{-1} + (20 k\Omega)^{-1})^{-1}}$ is the addition in parallel between the resistors, and $\mathrm{(18 C_{clk}\,\text{-{}-}\, C_v) = ((18 C_{clk})^{-1} + C_v^{-1})^{-1}}$ is the addition in parallel of the 18 $\mathrm{C_{clk}}$ in serie with $\mathrm{C_v}$. The area of the clock traces is 196\,mm$^2$, much larger than the video output maximum, 23.6\,mm$^2$. Therefore, $\mathrm{R_v >> R_{clk}/18}$, and $\mathrm{C_v << 18\,C_{clk}}$ so we can approximate $\mathrm{R_v + R_{clk}/18 \sim R_v}$ and $\mathrm{(18 C_{clk}\,\text{-{}-}\, C_v) \sim C_v}$. The signal amplitude is defined by \[\mathrm{\left|\frac{V_{out}}{V_{in}}\right| \propto \left|\frac{R_{out} \parallel 20 k \Omega}{Z_{sub} + R_{out} \parallel 20 k \Omega}\right|},\]

where $\mathrm{Z_{sub} = R_v \,\text{-{}-}\, C_v \,\text{-{}-}\, \frac{R_{clk}}{18} \,\text{-{}-}\, 18 C_{clk} \sim R_v \,\text{-{}-}\, C_v }$ is the impedance through the substrate.

As presented in Fig.~\ref{fig:signalSat}, the signal amplitude and $\tau$ increases with the trace length, L, which we can also infer from the definitions of $\,\mathrm{C_v}\,$ and $\,\mathrm{R_v}\,$: $\mathrm{C_v} \propto L$ while $\mathrm{R_v} \propto \frac{1}{L}$. Then, $\mathrm{\tau \sim (R_{out}\parallel20\,k\Omega)\,C_v+ R_v C_v}$ increases with $\mathrm{C_{v}}$, while the signal amplitude increases with $\mathrm{C_v}$ and decreases with $\mathrm{R_v}$.

To obtain $\mathrm{R_a}$, and in turn $\mathrm{R_{v,clk}}$, we used a CCD as in the multi-chip module but on a package that connects a single sensor directly to the readout electronics through a flex cable. We connected in parallel to $\mathrm{R_{out}}$ a square pulse generator, 2.2\,nF and 270\,pF capacitors that replicate $\mathrm{C_{clk}}$ and $\mathrm{C_v}$, and a variable resistor $\mathrm{R_{var}}$ tuned to obtain the same signal $\tau$ as in the data. We present this configuration in the bottom panel of Fig.~\ref{fig:substrate} (red-dashed box). Then \[\mathrm{R_{var} = R_v + R_{clk} = R_a \left(\frac{1}{A_v} + \frac{1}{A_{clk}}\right)},\]

with $\mathrm{R_{var} = 3.9\,k \Omega}$, $\mathrm{A_v = 23.6\,mm^2}$ (for the video output trace with 241\,mm), and $\mathrm{A_{clk} = 196\,mm^2}$, which gives $\mathrm{R_a = 82.15\,k \Omega\,mm^2}$. We then extract $\mathrm{R_{out}}$ using the recovery time of the signal, $\tau \sim 2.40$\,$\upmu$S in this case (Fig.~\ref{fig:signalSat} bottom), so $\mathrm{R_{out} \approx 7\,k\Omega}$, which is consistent with previous measurement of the transistor curves~\cite{Holland:2003}.

Finally, we computed the resistances associated with the aluminum lines using \[\mathrm{R} = \frac{\rho_\mathrm{Al}}{\sigma} \mathrm{L} = 0.25 \frac{\Omega}{\mathrm{mm}}\mathrm{L},\] where $\rho_\mathrm{Al}$ is the aluminum resistivity at 140\,K, $\sigma$ the cross-sectional area, and $L$ the trace length; these resistances, of $24\sim60\Omega$ for the video output and $\sim 1.5 \Omega$ for each clock, are negligible compared to the other contributions.

\begin{figure}[t]
    \centering
    \includegraphics[width=1\linewidth]{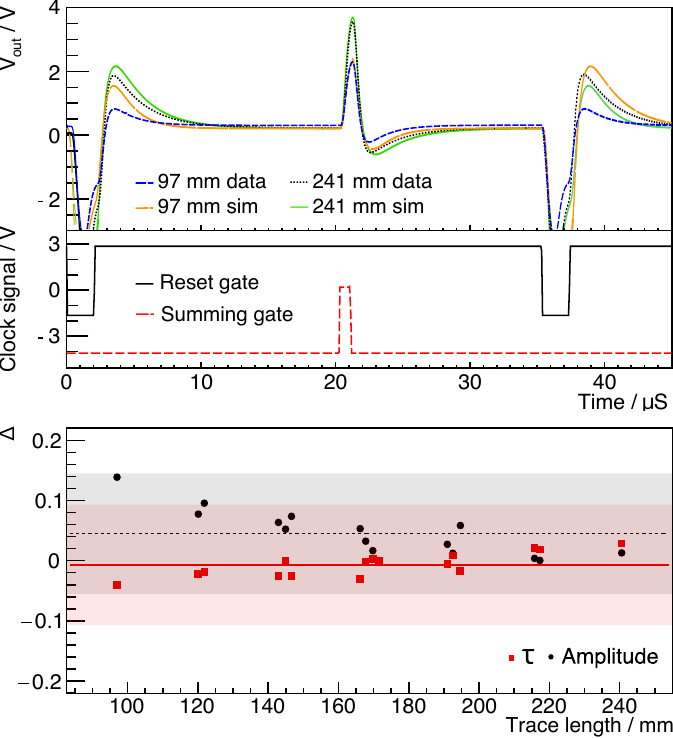}
    \caption{(Top) Video output signal as a function of time for traces with 97 and 241\,mm length. We show results for data and simulations on top, and the clocking for the rest and summing well gates at the bottom. (Bottom) Signal $\tau$ and Amplitude deviation ($\Delta$) between simulation and data as a function of the trace length.}
    \label{fig:signalSim}
\end{figure}

We implemented the signal model in an LTSpice\textsuperscript{\textregistered}~\cite{SPICE} simulation to understand the coupling for different configurations. To this end, we simulate clocking the summing well and reset gates to extract the coupling at the video output. In the top panel of Fig.~\ref{fig:signalSim}, we present the video output for 97\,mm and 241\,mm traces (top) as signal voltage in time. We show the simulation and measurements obtained during testing. Furthermore, we show the reset and summing well clocks (bottom). CCDs in this work have four output amplifiers with corresponding reset transistors, one on each corner. Although we read a whole CCD with one video output in the multi-chip module configuration, the remaining three are connected. Then, the clock amplifier drives 64 reset gates in parallel, producing a non-negligible capacitive load to the driver and slowing down the reset pulse falling and rising edges, effectively reducing the high-frequency component of the coupling from the reset gate. Since we did not consider this capacitive load in the simulation, we obtained a slight discrepancy between the simulation and data right after the reset gate, which does not affect the package performance.


To further validate the simulation, we show in the bottom panel of Fig.~\ref{fig:signalSim} the signal $\tau$ and amplitude fractional deviation between simulation and data, defined as $\mathrm{\Delta_{\tau} = \frac{\tau_{sim}-\tau_{data}}{\tau_{sim}}}$ (black circles) and $\mathrm{\Delta_{A} = \frac{A_{sim}-A_{data}}{A_{sim}}}$ (red squares),  as a function of the trace length. Horizontal lines represent the mean bias for $\tau$ (black dotted) and amplitude (solid red), while the gray and red shadow represent the 10\% deviation from the mean. In both cases, the simulation reproduces well the measurements with a maximum deviation of less than 15\%.

\section{Design optimization and production recipe}
\label{sec:newDesign}

Understanding the electrical behavior of the multi-chip module allows us to fine-tune the design and fabrication process to optimize the signal uniformity, suppress the coupling, and prevent electronics saturation. We implemented new designs that include a precision-length matching of the video output traces to 241\,mm and reduced trace areas by about half to increase $\mathrm{R_v}$ and decrease $\mathrm{C_v}$. Moreover, we incorporated a ground fill surrounding the traces to introduce a $\sim$\ 60\,nF decoupling capacitor to ground connected between $\mathrm{R_{clk}}$ and $\mathrm{R_v}$, as we illustrate in the right panel of Fig.~\ref{fig:substrate} (blue dashed box). This feature reduces the coupling by providing a low-impedance path from the substrate to ground for high-frequency signals.

To test the new design concepts, we fabricated two wafer flavors: one using the previous recipe, p$^{++}$-type silicon with 3\,$\mathrm{\upmu m}$ $\mathrm{SiO_2}$, and another with a 6\,$\mathrm{\upmu m}$ $\mathrm{SiO_2}$ layer on intrinsic silicon. The intrinsic silicon conductivity falls abruptly when cooled down, so $\mathrm{R_{clik,v} \rightarrow \infty}$,  strongly suppressing the substrate coupling. Plus, the thicker $\mathrm{SiO_2}$ layer drives down $\mathrm{C_{clk}}$ which mitigates the coupling even further. We also fabricated wafers with varying thicknesses of aluminum and gold to confirm that the resistivity of the metal line does not affect the coupling to the video outputs.

\begin{figure}[t]
    \centering
    \includegraphics[width=1\linewidth]{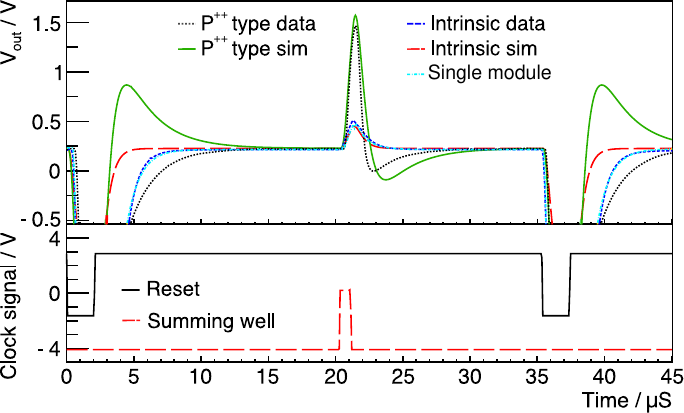}
    \caption{(Top) Video output signal as a function of time for p$^{++}$-type and intrinsic silicon substrates with the new design that includes the trace length matching to 241\,mm and ground fill. We present results for data and simulation, as well as a single module trace. (Bottom) Reset and summing well gates.}
    \label{fig:signalSimNew}
\end{figure}

We repeated the testing described in Section~\ref{sec:testing} for the wafers fabricated with the new designs. We present results for measurements and simulations of the video output (top) during the reset and summing well gates clocking (bottom) in Fig.~\ref{fig:signalSimNew}. Data show that the ground fill reduces the coupling amplitude for a trace length of 241\,mm in the p$^{++}$-type wafer from 4.2\,V to 1.5\,V with similar $\tau$, from $2.3\,\upmu$s to $2.1\,\upmu$s. Furthermore, the intrinsic silicon suppresses the coupling, eliminating the undershoot after the summing well and bringing the signal amplitude to 0.25\,V, identical to the ceramic substrate or the single module used in section~\ref{sec:sim}. As expected, we saw no differences in the coupling amplitude for p$^{++}$-type wafers with the old design and different aluminum thicknesses.

We verified that the output signal shape on the intrinsic silicon is similar to signals in the ceramic and single modules for different temperatures between 140\,K and 200\,K. We also performed several temperature cyclings and ran the module at about 140\ K for over a year to find no signs of mechanical stress.

\begin{figure}[t]
    \centering
    \includegraphics[width=1.0\linewidth]{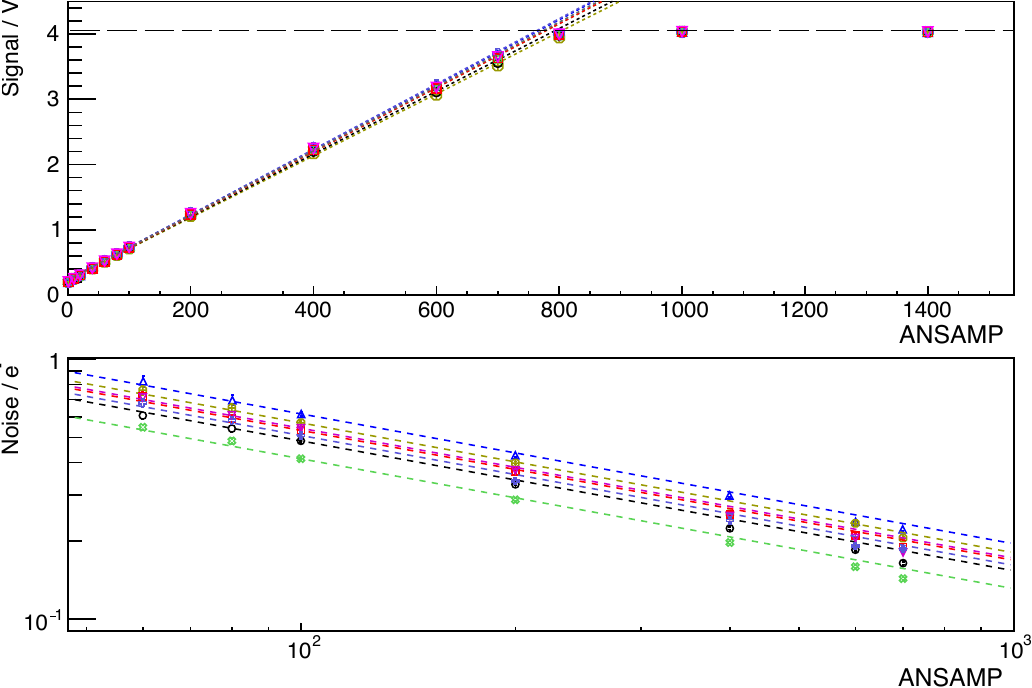}
    \caption{(Top) Signal output of empty pixels for different video channels in
the same multi-chip module fabricated on intrinsic silicon with the new design as a function of ANSAMP. (Bottom) noise as a function of ANSAMP for different channels with the $\frac{\sigma_1}{\sqrt{N}}$}
    \label{fig:noiseIntrinsic}
\end{figure}

To further validate the new design on intrinsic silicon, we present in the top panel of Fig.~\ref{fig:noiseIntrinsic} the mean empty pixel value as a function of the number of analog samples for different CCDs (different markers-colors) in the multi-chip module. Same as in Fig.~\ref{fig:signalSat}, dashed lines correspond to the extrapolation of a linear fit using the data obtained with ANSAMP < 100, and the black dashed line at about -4 V to the saturation point of the ADC. We selected the same channels, or CCD positions, as in Fig.~\ref{fig:signalSat} to confirm that the trace length matching ensures consistency across all video output signals. The slope corresponds to the fixed positive offset in the electronics reported in section~\ref{sec:results} and produces the saturation of all channels at about 800\,ANSAMP. 

In the bottom panel of Fig.~\ref{fig:noiseIntrinsic}, we present the readout noise as a function of the number of analog samples for the same channels as in the top panel. Dashed lines correspond to the expected noise as a function of the
number of samples: $\frac{\sigma_1}{\sqrt{N}}$, where we calculate $\sigma_1$ from the noise at 100 ANSAMP. The noise reduction behaves as expected until the electronics start to saturate. We specifically tuned the electronics to measure signals in a dynamic range from 0 to 10\,e$^-$ with a readout noise lower than 0.20\,e$^-$, which we demonstrated for 400\,ANSAMP~\cite{chavez2022_multiplexed}. Although these are OSCURA specifications, we could adjust the readout system to a different dynamic range if required.

In this manuscript, we report on four silicon multi-chip modules to outline the critical steps in optimizing the design and recipe for full-scale OSCURA production. Although we tested various fabrication and design iterations using different substrate types, metals, and SiO$_2$ layers, we omitted the results since the outcomes did not significantly impact the final design and recipes. For completeness, Table~\ref{table:wafers} summarizes the main wafer fabrication runs, highlighting their distinct characteristics in terms of fabrication process and design. Stars denote fully assembled and tested wafers, while we indicate the wafers explicitly reported in this work by bold letters, specifically, in section~\ref{sec:results}, we present results for Fabrication id. 1 and 3, in section~\ref{sec:sim} for fabrication id. 3, and in section~\ref{sec:newDesign} for fabrication ids. 8 and 10. It took about five iterations to achieve the first working N-type wafer (id. 1), resulting in a $\sim$20\% yield, and two attempts to produce the first P-type wafer (id. 3), with a $\sim$50\% yield. By the end of the development process, the yield for the final six wafers (ids. 5 to 10) reached 100\%. We attribute this significant improvement to the learning curve associated with mastering the process and optimizing the recipes over the iterations.

\begin{nolinenumbers}
\begin{table}[t!]
\setlength{\tabcolsep}{3pt}
\centering
\renewcommand{\arraystretch}{1.2}
\begin{tabular}{l l l l} 
\toprule
\textbf{Id.}  & \textbf{Silicon type, 1$^\mathrm{st}$ SiO$_2$} & \textbf{Metal, 2$^\mathrm{nd}$ SiO$_2$} & \textbf{Design} \\
\midrule
\textbf{1*} & \textbf{n, 390\,nm deposition} & \textbf{Al 450\,nm, 450\,nm} & \textbf{Original}  \\
2* & '' & '' & ''  \\
\textbf{3*} & \textbf{p$^{++}$, 3\,$\upmu$m thermal} & \textbf{''}  & \textbf{''}  \\
4 & '' & Gold 1\,$\upmu$m, none & '' \\
5* & '' & Al 2\,$\upmu$m, 2\,$\upmu$m &  '' \\
6* & '' & Al 1\,$\upmu$m, 1\,$\upmu$m & Trace length matching\\
\textbf{7*} & \textbf{''} & \textbf{Al 450\,nm, 450\,nm} & \textbf{Matching + ground fill} \\
8* & intrinsic, 6\,$\upmu$m thermal & '' & '' \\ 
\textbf{9*} & \textbf{''} & \textbf{''} & \textbf{Final design}\\
10* & '' & '' & ''\\
11 & '' & Gold 1\,$\upmu$m, none & Original \\

\bottomrule
\end{tabular}
\caption{Most relevant wafer fabrication runs according to the silicon substrate type, first and second SiO$_2$ layer process and thickness, metal type and thickness, and design. Stars (*) indicate fully assembled and tested wafers, while bold letters indicate the wafers we explicitly report in this article.}
\label{table:wafers}
\end{table}
\end{nolinenumbers}

\begin{figure}[t]
    \centering
    \includegraphics[width=0.9\linewidth]{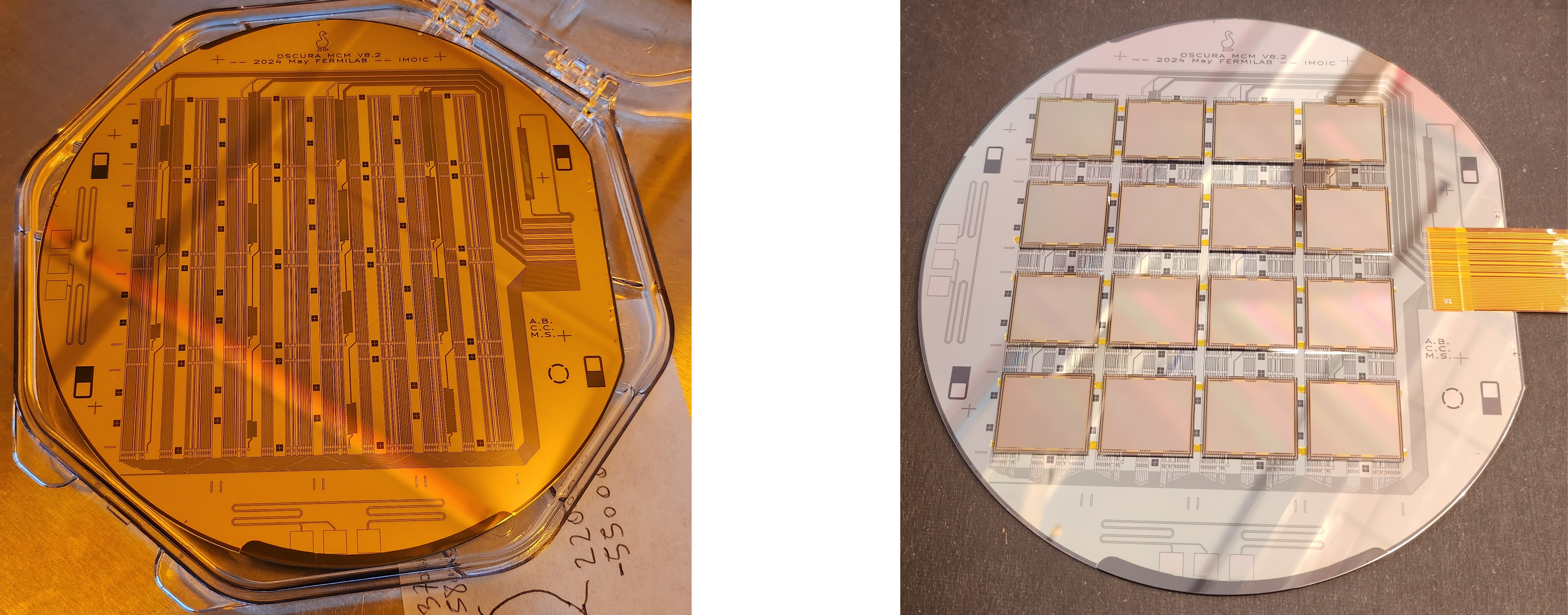}
    \caption{(Left) Final multi-chip module design fabricated on a intrinsic silicon substrate. The new design includes the length matching of the video output traces, a ground fill, and several markers and features for fabrication and assembly purposes. (Right) Same module after assembly with CCDs and flex cables attached.}
    \label{fig:finalMCM}
\end{figure}

We show a picture of a wafer with the ultimate multi-chip module design in the left panel of Fig.~\ref{fig:finalMCM}, featuring the trace length matching, about 50\,$\mathrm{\upmu}$m trace widths, and a ground fill fabricated on an intrinsic silicon wafer with a 6\,$\mathrm{\upmu}$m SiO$_2$ layer. We also present in the right panel of Fig.~\ref{fig:finalMCM} a picture of the module after assembly. These advancements cleared the way for the ongoing fabrication of 1500 wafers at Microchip Technology towards the whole 10\,kg OSCURA array.

Regarding radiogenic backgrounds, while the current facilities and equipment for evaluating radiation rates lack the sensitivity to assess the low-background levels required for OSCURA, there is, in principle, no fundamental obstacle to achieving this goal. These tests will be conducted in an ultra-low-background environment, which is not yet accessible.
The concept of silicon interconnects builds on the use of fabrication materials and processes similar to those used for CCDs, paired with analogous background mitigation strategies to address challenges such as radon plate-out~\cite{DAMICBGD} and cosmogenic activation~\cite{richardCosmo}. In this framework, meeting the background requirements for the multi-chip modules builds on the ongoing efforts to achieve the stringent fabrication standards established for the CCDs. Although no dedicated high-sensitivity study has yet been performed to assess the radiogenic background of the silicon interconnect, multiple measurements of OSCURA CCDs conducted at a shallow underground facility have not indicated any radiogenic concerns~\cite{oscurasensors}. Furthermore, tests on a silicon interconnect fabricated with similar materials and processes have shown that the package does not contribute to an increase in dark current~\cite{Sensei2020Sup,DAMICLBC}.

\section{Summary}
\label{sec:outlook}

In this work, we first presented the design, production, assembly, and testing of a novel silicon-based package for multi-kilogram skipper-CCD detectors, a.k.a. multi-chip modules. The collaboration between Argonne National Laboratory and Fermilab allowed us to produce the first functional silicon package with sub-electron noise. The tests performed to characterize its electrical properties showed a signal distortion associated with the coupling between the clock signals and the video outputs through the silicon substrate that depends on the area covered by the video trace, the Al and $\mathrm{SiO_2}$ thickness, the composition of the $\mathrm{Si-SiO_2}$ interface, and the material properties such as the substrate type. We observed that these electrical properties translate into saturation points due to the offsets introduced in each analog sample. 

We characterized the signals in a multi-chip module fabricated on p$^{++}$-type silicon with a $\mathrm{3\ \upmu m\ SiO_2}$ and $\mathrm{0.5\ \upmu m\ Al}$ layers, in terms of amplitude, offset, and recovery time to write an empirical signal model implemented in LTSpice\textsuperscript{\textregistered} and validated the simulations using laboratory measurements. The simulation described well the data with a maximum bias of less than 15\%. We then used the simulation to define adjustments in the design and fabrication to uniformize the video output signal and reduce the signal coupling.

We implemented a new design with trace length matching, reduced trace widths, and a ground fill, which we fabricated on p$^{++}$-type and intrinsic silicon. Measurements on both wafers show a significant reduction in the signal coupling, making the intrinsic silicon the most effective approach to suppress it. Therefore, we selected this new design on intrinsic silicon to proceed with the production of 1500 wafers at Microchip to implement 24000 CCDs, adding to a total of 10\, kg. We will conduct a detailed assessment of the dark current and radiogenic background of this package once a low-background vessel is ready in a deep underground environment.

\section{Acknowledgments}

This manuscript has been supported by Fermilab and authored by Fermi Research Alliance, LLC under Contract No. DE-AC02-07CH11359 with the U.S.~Department of Energy, Office of Science, Office of High Energy Physics, and the DOE Early Career Award (DOE-ECA). Work performed at the Center for Nanoscale Materials, a U.S. Department of Energy Office of Science User Facility, was supported by the U.S. DOE, Office of Basic Energy Sciences, under Contract No. DE-AC02-06CH11357. The CCD development work was supported in part by the Director, Office of Science, of the U.S. Department of Energy under No. DE-AC02-05CH11231. The United States Government retains and the publisher, by accepting the article for publication, acknowledges that the United States Government retains a non-exclusive, paid-up, irrevocable, worldwide license to publish or reproduce the published form of this manuscript, or allow others to do so, for United States Government purposes.

\bibliographystyle{ieeetran}
\bibliography{references}

\end{document}